\documentclass[%
 reprint,
superscriptaddress,
amsmath,amssymb, 
aps,
]{revtex4-2}
\usepackage{natbib}
\usepackage{lipsum} 
\usepackage{graphicx}
\usepackage{dcolumn}
\usepackage{bm}
\usepackage{mathtools}

\usepackage{hyperref}

\usepackage{lipsum} 
\usepackage{xcolor}

\newcommand{\ins}{\text{in}}
\newcommand{\out}{\text{out}}

\newcommand{\z}{\phi_0}

\newcommand{\io}{\text{in/out}}

\newcommand{\psicritpm}{\bar{\psi}_\text{crit}^\pm}
\newcommand{\psibinpm}{\bar{\psi}_\text{bin}^\pm}

\newcommand{\eq}[1]{Eq.~(\ref{eq:#1})}
\newcommand{\fig}[1]{Fig.~\ref{fig:#1}}

\newcommand{\qd}{\, .}
\newcommand{\qc}{\, ,}

\newcommand{\vect}[1]{\boldsymbol{#1}}

\begin{document}


\title{
Critical transition
between intensive and extensive active droplets
}


\author{Jonathan Bauermann
}
\affiliation{Department of Physics, Harvard University, Cambridge, MA 02138, USA}

\author{Giacomo Bartolucci
}
\affiliation{Department of Physics Universitat de Barcelona, Carrer de Martí i Franquès 1-11, 08028 Barcelona, Spain}

\author{Job Boekhoven
}
\affiliation{
School of Natural Sciences, Department of Bioscience,
Technical University of Munich, Lichtenbergstraße 4, 85748 Garching, Germany}

\author{Frank Jülicher}
\affiliation{
Max Planck Institute for the Physics of Complex Systems,
 Nöthnitzer Stra\ss e~38, 01187 Dresden, Germany
}
\affiliation{
Center for Systems Biology Dresden,  Pfotenhauerstra\ss e~108, 01307 Dresden, Germany
}
\affiliation{
Cluster of Excellence Physics of Life, TU Dresden, 01062 Dresden, Germany 
}

\author{Christoph A. Weber}
\affiliation{
 Faculty of Mathematics, Natural Sciences, and Materials Engineering: Institute of Physics, University of Augsburg, Universit\"atsstra\ss e~1, 86159 Augsburg, Germany
}



\date{\today}

\begin{abstract}
Emulsions ripen with an average droplet size
increasing in time.
In chemically active emulsions, 
coarsening can be absent, leading to a non-equilibrium steady state with mono-disperse droplet sizes. 
By considering a  minimal model for phase separation and chemical reactions maintained away from equilibrium, we show that there is a critical transition in the conserved quantity between two classes of chemically active droplets:
intensive and extensive ones.
Single intensive active droplets reach a stationary size mainly controlled by the reaction-diffusion length scales.
Intensive droplets in an emulsion interact only weakly, and the stationary size of a single droplet approximately sets the size of each droplet.
On the contrary, the size of a single extensive active droplet scales with the system size, similar to passive phases. In an emulsion of many extensive droplets, their sizes become stationary only due to interactions among them. 
We discuss how the critical transition between intensive and extensive active  droplets affects shape instabilities, including the division of active droplets, paving the way for the observation of successive division events in chemically active emulsions. 
\end{abstract}

\maketitle

\section{Introduction}

Coarsening or ripening refers to the growth of larger domains at the expense of smaller domains that eventually shrink. For passive systems, the kinetics of coarsening stops when the system reaches  equilibrium, corresponding to a single domain in a finite system. Coarsening occurs in various systems, ranging from spin systems~\cite{williamson2016universal}, liquid emulsions~\cite{bibette1999emulsions}, and crystallized precipitates~\cite{pollock1994directional}. The kinetics of coarsening  is universal and determined by conservation laws, symmetries, and the dimension of the system~\cite{wagner1961theorie,lifshitz1961kinetics, voorhees1985theory, Krichevsky1993}. 

For active systems persistently maintained away from equilibrium~\cite{marchetti2013hydrodynamics,doostmohammadi2018active, shankar2022topological, weber2019physics}, the kinetics of coarsening is altered and can be even suppressed~\cite{glotzer_monte_1994, zwicker_suppression_2015, tjhung2018cluster, Brauns2021, weyerCoarseningWavelengthSelection2023}. The paradigm is reaction-diffusion systems that give rise to non-equilibrium steady state patterns with various spatial morphologies~\cite{langer1980instabilities, cross1993pattern, frey2018protein}. Another example is Model B+, which, in contrast to the classical Model B, also accounts for contributions to the diffusive fluxes that do not arise from free energy~\cite{wittkowski2014scalar, cates2022active}. These fluxes give rise to anti-coarsening with a condensed phase that stopped growing and a ``bubbly”  morphology of material-poor domains~\cite{tjhung2018cluster}. Finally, suppressed ripening was also observed in liquid-liquid phase-separated systems with chemical reactions maintained away from equilibrium~\cite{glotzer_monte_1994, zwicker_suppression_2015, wurtz2018chemical, demarchiEnzymeEnrichedCondensatesShow2023}. These systems are also called chemically active emulsions~\cite{weber2019physics, zwicker2024chemicallyactivedroplets}.  

The formation of steady-state patterns in reaction-diffusion systems relies on the reaction flux that breaks the detailed balance of the rates. Together with diffusion, this gives rise to various reaction-diffusion length scales that are crucial but not exclusively responsible for pattern morphology. Chemical processes generically come with conservation laws for mass, and if incompressible, also for volume. It has been shown that conservation laws are key determinants for the emerging patterns in mass-conserving reaction-diffusion systems~\cite{halatek2018rethinking, burkart2022control, weyerCoarseningWavelengthSelection2023}. Key implications are that the pattern-forming transition is typically sub-critical, reminiscent of discontinuous phase transitions~\cite{halatek2018rethinking}. 

Active emulsions also give rise to reaction-diffusion length scales. These length scales are crucial for various non-equilibrium phenomena in active emulsions, such as dividing droplets~\cite{zwicker_growth_2017, seyboldt2018role, bauermann_energy_2022}, formation of steady liquid shells~\cite{bartolucci2021controlling, bergmann_liquid_2023, bauermann_formation_2023}, and the suppression of Ostwald-ripening~\cite{zwicker_suppression_2015, sastre2024size}. Despite such interesting phenomena, a limitation of some minimal models~\cite{glotzer_monte_1994, zwicker_suppression_2015} is that two components were considered lacking  conservation laws. Similar to reaction-diffusion systems~\cite{halatek2018rethinking}, conservation laws can qualitatively alter the nature of the transition and instabilities in active emulsions. 

In passive, phase-separated systems  with chemical reactions, the reaction-diffusion length scales do not determine the equilibrium state.
In this passive case, conservation laws, i.e., so-called lever rules for quantities conserved by the chemical reactions, determine the volume of the condensed phase(s) at thermodynamic equilibrium. When the chemical reactions are maintained away from equilibrium (active emulsion), the emerging reaction-diffusion length scales compete with the conservation laws. It remains unclear whether such reaction-diffusion scales or the conservation law dominate the selection of pattern length scales in active emulsions.

\begin{figure}
\centering
\includegraphics[width=0.48\textwidth]{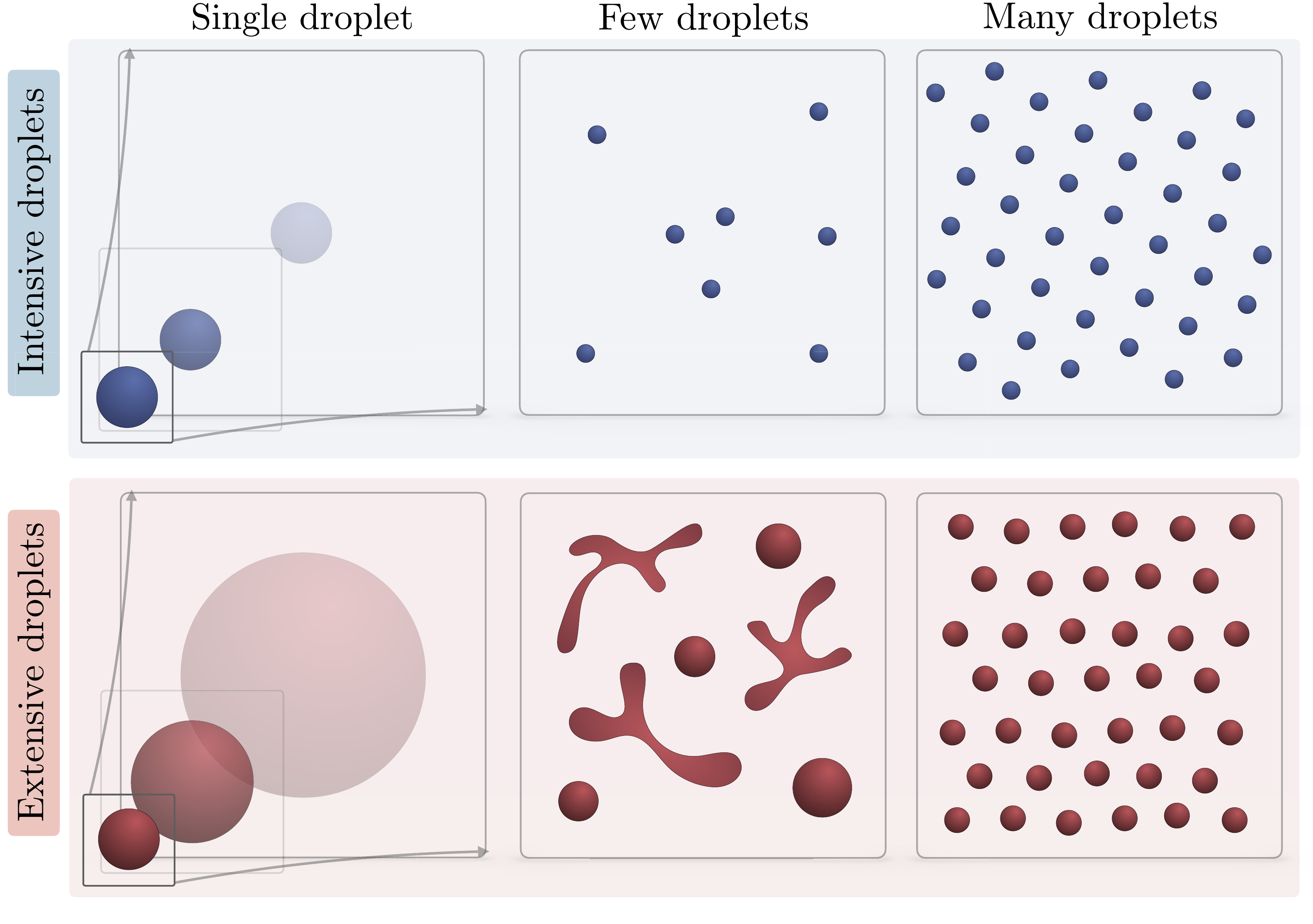}
\caption{\textbf{Intensive and extensive active droplets and their behavior in emulsions.}
Chemically active emulsions undergo a transition between intensive  (blue) and extensive (red) droplets.
The transition depends on the value of the quantity conserved by the chemical reaction (Eq.~\eqref{eq:cons_quantity}).
Intensive droplets (upper row) have finite radii and interact mildly when many droplets are present in an emulsion.
On the contrary, extensive droplets (bottom row) grow until interactions between them arrest the growth.
\label{fig:sketch}}
\end{figure}

Here, we study the role of conservation laws in multi-component mixtures for the droplet size and the droplet number in monodisperse active emulsions that do not undergo coarsening.  
Our key finding is that the conserved quantity controls a critical transition between 
intensive and extensive chemically active droplets that differ in their physical behavior when increasing the system size; see Fig.~\ref{fig:sketch} for an illustration. In the case of intensive droplets, single droplets in large systems are stationary, with droplet sizes independent of system size. 
For extensive droplets, the stationary droplet size increases with the system size. 
We study the consequences of this transition for the collective dynamics of many droplets in emulsions and show how monodispersity can arise for both classes of chemically active droplets.

\section{Minimal model for an active emulsion with a conserved quantity}

\subsection{Dynamics of the concentration fields}

An incompressible binary mixture with two molecules $A$ and $B$ converting into each other via a chemical reaction has a conserved quantity, the total mass of $A$ and $B$. However, in a binary mixture, this conserved quantity is trivial as it is constant in space and thus does not change dynamically~\cite{glotzer_monte_1994, zwicker_suppression_2015}. 
A mixture containing chemical reactions must comprise at least three different molecules to have a conserved quantity that can vary in space. 
In the following, we introduce a minimal model for such a  ternary mixture that is incompressible and composed of a non-reacting solvent $S$ and the molecules of types $A$, $B$, that can react via the following reaction scheme
\begin{equation}
\label{eq:reaction_scheme}
A \xrightharpoonup{k_{BA}} B\, , \quad 
B \xrightharpoonup{k_{AB}} A
\qd 
\end{equation}
For this chemical reaction,
the dynamics of average concentrations $\bar{\phi}_i(t)=V^{-1}\int_V d^3x \, \phi_i(\vect{x},t)$ ($i=A,B,S$) are constraint to a specific conserved line in the $\bar{\phi}_A$-$\bar{\phi}_B$-plane which can be expressed as
\begin{equation}\label{eq:cons_quantity}
    \bar{\psi}  =  (\bar{\phi}_A+\bar{\phi}_B)/2 \qc
\end{equation}
where $V$ denotes the volume of the system and $\phi_i(\vect{x},t)$ are the concentration fields  that depend on position $\vect{x}$ and time $t$. The conserved quantity is denoted as $\bar{\psi}$.

The dynamical equations for the fields $\phi_A(\vect{x},t)$ and $\phi_B(\vect{x},t)$ are
\begin{subequations} \label{eq:dyn_phi} 
\begin{align}
\partial_t \phi_A &=   \nabla \cdot \left( \Gamma_A \nabla \mu_A \right)+ k_{AB} \phi_B  - k_{BA}\phi_A  \qc \label{eq:dyn_phiA}\\ 
    \partial_t \phi_B &=  \nabla \cdot \left( \Gamma_B \nabla \mu_B \right) - k_{AB} \phi_B  + k_{BA}\phi_A  \qc \label{eq:dyn_phiB}
\end{align} 
\end{subequations}
where $\mu_i =  \delta F /\delta \phi_i$
are the chemical potentials of $i=A,B$, descending from the Helmholtz free energy $F=\int_V d^3x \left( f_0+ \kappa_A (\nabla \phi_A)^2/2 + \kappa_B (\nabla \phi_B)^2/2 \right)$ \cite{Hohenberg1977}. Here,  $\Gamma_i$ denote the mobility coefficients.
The first terms of Eqs.~\eqref{eq:dyn_phi}  are diffusive fluxes that are driven by local gradients of the chemical potentials, while the remaining terms, $k_{AB} \phi_B  - k_{BA}\phi_A$, are chemical fluxes. For passive systems, 
the reaction rate coefficients $k_{ij}$ are concentration dependent such that the free energy $F$  determines  thermodynamic equilibrium~\cite{bauermann_energy_2022, bauermann_chemical_2022}. If, however, $k_{ij}$ are also dependent on some external reservoir or are chosen independently of $F$, the system is inherently chemically-active~\cite{
glotzer_monte_1994, Lefever_comment, Glotzer_reply,
zwicker_suppression_2015, weber2019physics,
bauermann_chemical_2022}.

\begin{figure}
\centering
\includegraphics[width=0.49\textwidth]{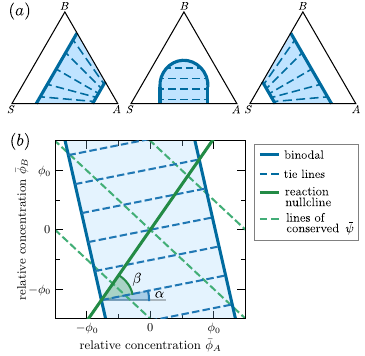}
\caption{\textbf{Geometrical representation of phase diagram and reaction nullcline}.
\textbf{(a)} Sketches of phase diagrams of different ternary mixtures: In the white domains, the system is homogeneous, while in the blue domains, it is phase-separated. Each diagram corresponds to different interactions among the molecules $A$, $B$, and $S$. 
The key qualitative properties of such diagrams can be captured by a  geometrical representation.
\textbf{(b)} Geometrical representation of a phase diagram with the
binodal line (dark solid blue), 
the tie-lines (dashed blue),
the conserved quantities $\bar \psi$ (dashed green), 
and the reaction nullcline (solid green), where chemical fluxes vanish. 
Moreover, $\alpha$ is the \textit{compositional angle} determining the composition of the coexisting phases, while  $\beta$ is the
\textit{activity parameter} characterizing the strength of non-equilibrium driving. Only for $\beta \not= 0$, 
there can be non-equilibrium steady states with phase compositions that are not connected by a tie line. 
As a result, there are chemical and diffusive fluxes in the non-equilibrium steady state.
}
\label{fig:ph_d}
\end{figure}

Without chemical reactions, the dynamic system governed by Eqs.~\eqref{eq:dyn_phi} relaxes to phase equilibrium, characterized by equal chemical potentials $\mu_i^\text{I}=\mu_i^\text{II}$, and equal osmotic pressures 
$\Pi^\text{I}=\Pi^\text{II}$, where $\Pi= -f+ \sum_{i=A,B} \phi_i \mu_i$, see Ref.~\cite{safranStatisticalThermodynamicsSurfaces2019a, kardarStatisticalPhysicsFields2007} for a general introduction. 
Such phase equilibria are also relevant for the dynamic system with chemical reactions since
they locally govern the dynamics of the interface.  
The phase equilibria of the mixture are portrayed by the phase diagram. 
In \fig{ph_d}(a), we show sketches for three different mixtures where
$A$ always phase separates from $S$ when $B$ is absent. 
The molecular interactions of $B$ with $A$ and $S$, shape the phase diagram. 
In the left panel of \fig{ph_d}(a), we show a sketch of a phase diagram where $B$ molecules are similar to $S$ and phase separate from $A$; in the central panel, $B$ molecules that are neutral and do not interact differently with  $A$ or $S$. In the right panel $B$ molecules are identical to $A$ molecules and therefore phase separate from the solvent. 

To mimic these interaction characteristics and to avoid tying our conclusions to specific free energy models, in the following, we take a geometrical view on phase diagrams.
Phase equilibria are characterized by the slope of the tie lines in the phase diagram. 
The relevant geometrical property of the phase diagrams that we consider is the local angle $\alpha$ between the tie lines formed with $\phi_A$-axes of the phase diagram (\fig{ph_d}(b)).
This local, geometrical view on a phase diagram allows us to consider free energies that are mathematically simpler even than mean field theories such as the Flory-Huggins theory, allowing us to obtain analytical results for stationary droplet states.
For ease of presentation, we consider cases where tie lines are locally orthogonal to both the dense and dilute branches of the binodal line. 
These geometrical properties can be captured by introducing the shifted concentration field $\phi_i \to \phi_i-\bar{\phi}_i$ and the following Ginzburg-Landau-like free energy density $f_0$:
\begin{subequations}\label{eq:f}
\begin{align}
f_0(\phi_A, \phi_B;\alpha) &= \frac{b_1}{2} (\phi_1 + \z )^2 (\phi_1 - \z )^2+ \frac{b_2}{2} \phi_2^2 \; ,\\
\label{eq:rot}
\begin{pmatrix}
\phi_1\\
\phi_2
\end{pmatrix} 
&=
\begin{pmatrix}
\cos \alpha & \sin \alpha\\
-\sin \alpha & \cos \alpha
\end{pmatrix} 
\begin{pmatrix}
\phi_A\\
\phi_B
\end{pmatrix}  \qc
\end{align}
\end{subequations}
where $b_1$ and $b_2$ are parameters characterizing interactions and entropic contributions. 
The qualitative cases shown in \fig{ph_d}(a) can be described 
by varying the angle $\alpha$ that parameterizes the rotations of the energy density $f_0$.
Specifically, different values of $\alpha$ correspond to different  types of interactions of the $B$ molecules with $A$ and $S$.  
Without loss of generality, we restrict ourselves to the domain $-\pi/4 \leq \alpha \leq \pi/4$. Indeed, due to the symmetry of the free energy, the transformation $\alpha' = \alpha + \pi/2$ is equivalent to re-labeling $A $ to $ B$ and $B $ to $ A$. 
For $\alpha = -\pi/4$, $B$ and  $A$ interact equally with the solvent, for $\alpha = 0$, $B$ does not interact with $A$ and solvent, and for $\alpha = \pi/4$, $B$ and $S$ interact equally with $A$. 
As a consequence, in the case $\alpha = -\pi/4$, the two reactants $A$ and $B$ localize in two distinct phases, i.e., they segregate.  In contrast, for $\alpha = +\pi/4$, molecules of type $A$ and $B$   phase separate together from the solvent $S$. Following Refs.~\cite{Deviri2021,Zeng2023}, we refer to $\alpha = -\pi/4$ as the \textit{segregative} case and $\alpha = +\pi/4$ as \textit{associative} case. In the literature on mixtures composed of a large number of components, these two cases are often called {demixing} and {condensation}, respectively~\cite{Thewes2023}. 
In the literature of biomolecular condensates, the shape of the phase diagrams is often related to the underlying molecular interactions.
According to Ref.~\cite{Riback2020}, {heterotypic} interactions correspond to the {segregative} case and {homotypic} interactions to the associative case.
Since the molecular interactions determine the composition of the coexisting phases,  we refer to  $\alpha$ as \textit{compositional angle}.
Varying $\alpha$ in the range $-\pi/4<\alpha<\pi/4$  interpolates  between the limits of the segregative and associative case.

In \fig{ph_d}(b), we show the phase diagram associated with the free energy density $f_0$ in \eq{f}. 
Note that the binodal lines (dark blue) separating the mixing from the demixing regimes (white and blue shaded area, respectively) are straight lines tilted by the compositional angle $\alpha$ from the vertical. 
The tie lines (represented as dashed blue lines) are parallel to each other, perpendicular to the binodal, and thus tilted by the compositional angle $\alpha$ from the horizontal.
Finally,  we comment on the role of $\phi_0$ in \eq{f} (\fig{ph_d}(b)): 
the value $\phi_0$ sets the scale of the concentration axis since for $\cos(\alpha)\phi_A + \sin(\alpha) \phi_B$ inside the interval $[-\phi_0, \phi_0]$, the passive system can phase separate.

In the following, we consider constant chemical rate coefficients $k_{ij}$ that are independent of the free energy $F$. As a result, the system is chemically active since in general no phase equilibrium of coexisting phases exists where the chemical reactions are at a steady state in all phases. 
These homogeneous steady states of chemical reactions are governed by the condition 
\begin{equation}\label{eq:reaction_nullcline}
k_{AB} \phi_B  = k_{BA}\phi_A 
\end{equation}
 defining the \textit{reaction nullcline} shown as a solid green line in \fig{ph_d}(b).
We refer to the angle $\beta$ between tie lines and the reaction nullcline with
\begin{equation}\label{eq:beta_def}
	\beta = \arctan(k_{BA}/k_{AB}) - \alpha \qc
\end{equation}
 as the \textit{activity parameter}. This is because it
 characterizes the strength of non-equilibrium driving.
Indeed, only for the special case $\beta = 0$, 
the system is passive and can settle to thermodynamic equilibrium because the coexisting concentrations (phase equilibrium) are also connected by the 
reaction nullcline (Eq.~\eqref{eq:reaction_nullcline}). Therefore, for $\beta=0$, the phase equilibrium and chemical equilibrium coincide.  
For $\beta \neq 0$, coexisting concentrations connected by a tie line do not lie on the reaction nullcline. 
Thus, no stationary state of reactions (\eq{reaction_nullcline}) simultaneously fulfils phase equilibrium. Instead, reaction fluxes and diffusive flux between the phases balance each other leading to a non-equilibrium steady state. 
The activity parameter is restricted to the domain $0 \leq \beta < \pi/4$ because for $\beta = \pi/4$ the reaction nullcline becomes parallel to the tie lines.

In summary, the active emulsions with the chemical reaction~\eqref{eq:reaction_scheme} 
is characterized, when using a geometrical representation of the phase diagram (\eq{f}), by the 
compositional angle $\alpha$, the activity parameter $\beta$, and the conserved quantity $\bar{\psi}$ (Eq.~\eqref{eq:cons_quantity}).

\subsection{Dynamics of the conserved and non-conserved fields}
\label{sec:dyn_con_noncon}

To understand the role of conserved quantities in the ripening kinetics, we introduce the conserved field $\psi(\vect{x},t)$ and the reaction extent field $\xi(\vect{x},t)$, which are defined as: 
\begin{subequations}\
\begin{align}
    \psi(\vect{x},t)&=(\phi_A(\vect{x},t) + \phi_B(\vect{x},t))/2 \qc \\
    \xi(\vect{x},t)&= (\phi_A(\vect{x},t) - \phi_B(\vect{x},t))/2 \qd
\end{align}
\end{subequations}
For the Ginzburg-Landau type of free energy 
(\eq{f}), we consider a constant mobility for simplicity and also write $\Gamma_i = \Gamma$, leading to: 
\begin{subequations}
 \label{eq:dyn_psi_xi} 
\begin{align}
\partial_t \psi &= \Gamma\nabla^2 \mu_\psi  (\psi,\xi; \alpha)   \qc \label{eq:dyn_psi}   \\
\partial_t \xi &= \Gamma \nabla^2 \mu_\xi  (\psi,\xi; \alpha) 
 \label{eq:dyn_xi} 
 \\
\nonumber
&\quad -K \xi + K \cot\left(\frac{\pi}{4} +\alpha+\beta\right) \psi \qc
\end{align} 
\end{subequations}
where we have introduced the chemical potentials of the conserved quantity, $\mu_\psi =(\mu_A +\mu_B )/2$, and the one of the non-conserved quantity, $\mu_\xi = (\mu_A-\mu_B)/2$. 
When recasting the free energy $f_0(\psi,\xi)$ in terms of the conserved and non-conserved fields, these chemical potentials can be obtained from functional derivatives, $\mu_\psi=\delta F/\delta \psi$ and $\mu_\xi=\delta F/\delta \xi$.
Furthermore, we define the overall reaction rate as
\begin{equation}\label{eq:overall_reaction_rate}
    K= k_{AB} + k_{BA} \qd
\end{equation}
In Appendix~\ref{app:chem_rates}, we show how 
the rate coefficients $k_{AB}$ and $k_{BA}$ can be expressed in terms of the overall rate $K$, the activity parameter $\beta$ and the compositional angle $\alpha$.

We can identify two special cases where the chemical potentials of the conserved field and of the reaction extent field decouple, i.e., the free energy density is devoid of coupling terms proportional to $(\psi \, \xi)$: 
(i) For the segregative case  $\alpha=-\pi/4$, we find $f_0(\psi,\xi) =  b_2 \psi^2/4 + b_1(\xi-\sqrt{2}\phi_0)^2(\xi+\sqrt{2}\phi_0)^2/8$. Thus, the free energy density contribution in $\xi$ is a classical double-well potential, which leads to a Cahn-Hilliard dynamics with reactions for the reaction extent field $\xi(\vect x, t)$ (\eq{dyn_xi}). The quadratic term of $\psi$ to the free energy density gives rise to Fick's law of diffusion for the conserved field $\psi(\vect x,t)$ ((\eq{dyn_psi})).
(ii) On the contrary, in the associative case, $\alpha=\pi/4$ gives a free energy density   $f_0(\psi,\xi) = b_1(\psi-\sqrt{2}\phi_0)^2(\psi+\sqrt{2}\phi_0)^2/8 + b_2 \xi^2/4$. In this case, the free energy density leads to a Cahn-Hilliard dynamics for the conserved field $\psi(\vect x,t)$ (\eq{dyn_psi}),  while the reaction extent field $\xi(\vect x, t)$ follows Fick's law of diffusion with linear reactions ((\eq{dyn_xi})). 
Though the dynamics of the conserved field $\psi$ evolves independently from the reaction extent field in both cases (i) and (ii), 
the dynamics of the reaction extent is affected by the conserved field $\psi(\vect x,t)$ via a source or a sink term in the reaction dynamics (\eq{dyn_xi}).

The ripening dynamics of these two special cases $\alpha=\pm \pi/4$ is known and has been excessively studied in the past:  
(i) For the segregative case $\alpha = -\pi/4$ ($A$ and $B$ segregate in different phases), the conserved variable $\psi$ follows a simple diffusion dynamics that is agnostic to the chemical reactions and the dynamics of the reaction extent $\xi$.
Thus, the field $\psi$ settles in a homogeneous state  on long-time scales. 
Moreover, the dynamics of $\xi$ is the dynamics of a binary mixture composed of $A$ and $B$  with active chemical reactions between these two components. This leads to the suppression of Ostwald ripening, i.e., a mono-disperse emulsion with a stationary droplet size~\cite{glotzer_monte_1994, zwicker_suppression_2015, wurtz2018chemical}.
Once droplet size exceeds this stationary value,  
droplets shrink or can undergo shape instabilities, giving rise to droplet division~\cite{zwicker_growth_2017}, or form stationary, active shells~\cite{bergmann_liquid_2023, bauermann_formation_2023}.
(ii) For the associative case $\alpha = \pi/4$ ($A$ and $B$ phase separate together), the conserved variable $\psi$ follows a classical Cahn-Hilliard dynamics. Thus, droplet-like domains, e.g., rich in $\psi$, undergo Ostwald ripening in a sea of low $\psi$, where bigger droplets grow at the expense of smaller shrinking ones that eventually disappear.
On large time scales and in a finite system, the system evolves toward a single droplet with a volume that scales with the system size $V$.  
The dynamics of the reaction extent $\xi$ 
is affected by the $\psi$ field without disturbing it. 
A similar decoupling occurs in models for reacting diluted ``client'' particles corresponding to $\xi$ and phase separating scaffold components corresponding to $\psi$~\cite{laha2024chemical}.

\begin{figure*}
\centering
\includegraphics[width=0.98\textwidth]{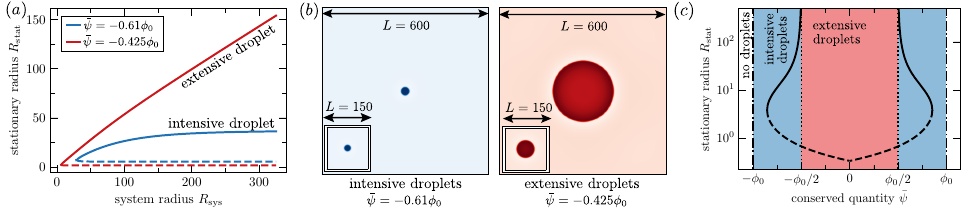}
\caption{ \label{fig:system_size} \textbf{Critical transition between intensive (blue) and extensive (red) active droplets.
}
\textbf{(a)} There are two classes of stationary solutions, depending on the value of the conserved quantity $\psi$.
The stationary droplet size $R_\text{sys}$ either increases with system size $R_\text{sys}$ (red line), or saturates for large values of $R_\text{sys}$ (blue line). 
We refer to theses cases as \textit{extensive} (red) and \textit{intensive} (blue) active droplets. 
Stable branches are depicted by solid lines and unstable branches are dashed. 
\textbf{(b)} 
Stationary concentration fields corresponding to each class (color code as in (a)) are shown for large ($L=600$) and small systems ($L=150$). They are obtained  from numerically solving Eq.~\eqref{eq:dyn_phi}. 
Note that the large red droplet would undergo a shape instability once the spherical droplet shape is perturbed.
\textbf{(c)} Dependence of the stable (solid line) and unstable (dashed line) stationary radius, $R_\text{stat}$, on the conserved quantity $\bar{\psi}$ for $\alpha = 0$ in an infinite system.
The stable branch (solid) diverges following a power law,  
$R_\text{stat} \propto |\bar{\psi} - \bar{\psi}_\text{crit}^\pm|^{-1}$, for $\bar{\psi} \gtrless \bar{\psi}_\text{crit}^\pm$, supporting that the transition in the conserved quantity is critical. 
Furthermore, we show the binodal (dashed-dotted line) and the $\psi$-value for the critical  transition between  (dotted line). 
Due to our choice of the phase diagram, the system is symmetric around $\bar{\psi}=0$.
Note that $\bar{\psi}<0$ corresponds to an $A$-rich droplet surrounded by a $S$-rich phase, while $\bar{\psi}>0$ corresponds to an $S$-rich droplet surrounded by an $A$-rich phase. 
We color-coded the regions where no droplets (white), intensive active droplet (blue), and extensive active droplet (red) exists for a system for large $K$, such that stationary radii get large and Laplace pressure effects can be neglected.
Details on the analytical calculation of $R_\text{stat}$
are given in Appendix~\ref{app:psi_crit}, and  parameter choices for all three panels are discussed in Appendix~\ref{sect:parameters_methods}.
}
\end{figure*}

In the remaining general cases $-\pi/4<\alpha<\pi/4$, the dynamics of the conserved field $\psi$ and the reaction extent field $\xi$ are coupled in the free energy. 
The key property as compared to the special cases (i,ii) is that the dynamics of the conserved field $\psi$ is influenced by the non-conserved field $\xi$. This coupling is a generic consequence of the different interactions among the molecular constituents of the mixture (see \fig{ph_d}(a,b) and related discussions). 
Therefore, the active  chemical reactions in the dynamic equation for the reaction extent $\xi$ (\eq{dyn_xi}) also affect the dynamics of the conserved field $\psi$ (\eq{dyn_psi}). 
In the next section, we will show that, strikingly, this remains relevant in the thermodynamic limit of large systems ($V \to \infty$), i.e., on length scales much larger than the reaction-diffusion length scale. 
Note that solely the non-conserved field $\xi$ gives rise to a reaction-diffusion length scale $\lambda = \sqrt{D/K}$, which can be obtained when linearizing near phase equilibrium. Here, $D = \Gamma \partial^2 f_0(\psi^\pm, \xi^\pm)^2/ \partial \xi^2$ which can be calculated in our model:
$D = \sqrt{(\Gamma \left[ b_2+ 4 b_1\phi_0^2 +(b_2- 4 b_1\phi_0^2)\sin(2\alpha)\right])/4}$.
The reaction-diffusion length scale $\lambda$ is finite and real (for $\Gamma>0$, $K >0$) for all values $-\pi/4 \leq \alpha \leq \pi/4$ and independent of $\beta$ and the conserved quantity $\bar \psi$.

\section{Stationary single droplets}\label{sect:single_drop}

To study the effects of the conserved quantity  $\bar{\psi}$ on the droplet size, we first consider a single spherical droplet in a finite, spherically-symmetric system of radius $R_\text{sys}=( 3V/(4\pi))^{1/3}$, where $V$ is the system volume. 
We consider a sharp interface limit, which is  valid when the droplet size is large compared t exceeds the width of the interface~\cite{brayTheoryPhaseorderingKinetics1994}.
Phase equilibrium is imposed at the position of this sharp interface. This boundary condition couples the reaction-diffusion equations in each phase. These  reaction-diffusion equations can linearized near phase equilibrium at the interface;  for details, see Appendix~\ref{app:edm} and Ref.~\cite{bauermann_energy_2022}. 

In the following, we discuss the stationary solutions for a single droplet, i.e., $\partial_t \psi(\vect{x},t) =0$ and $\partial_t \xi(\vect{x},t)=0$, and a stationary interface position $R_\text{stat}$. 
For large enough system sizes $R_\text{sys}$, we always find two solutions for $R_\text{stat}$ (solid and dashed lines in \fig{system_size}(a)) at which the total reaction flux of every component in one phase is perfectly balanced by the diffusive flux of the same component over the interface. 
The lower branch (dashed) of these two solutions is the critical nucleation radius corresponding to an unstable fixed point.   The upper branch (solid) of $R_\text{stat}$ is a stable fixed point corresponding to a non-equilibrium steady state. 
See Appendix~\ref{app:edm}, \eq{analytics_Rstat}, for an analytical expression of the stationary radius as a function of the activity parameter $\beta$ and the compositional angle $\alpha$ for large system sizes.

\begin{figure*}
\centering
\includegraphics[width=0.98\textwidth]{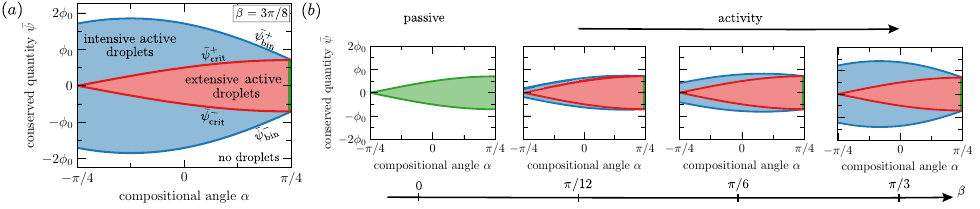}
\caption{\label{fig:state_diag} 
\textbf{State diagrams for intensive (blue) and extensive (red) active droplets.} 
Here we show for which values of the activity parameter $\beta$ (\eq{beta_def}), the compositional angle $\alpha$ (\eq{f}), and the conserved quantity $\bar \psi$ (\eq{cons_quantity}), we find intensive active droplets (blue) or extensive active droplets (red). %
\textbf{(a)} For an intermediate value of the activity parameter $\beta=3\pi/8$, we find two regimes: intensive active droplets (blue) and extensive active droplets (red).
For larger compositional angles $\alpha$, extensive droplets get favored. For the associative case $\alpha=\pi/4$, droplets behave like in the passive case (green line) (see Sec.~\ref{sect:arrest_ripe}).
\textbf{(b)} For an active parameter of $\beta=0$, passive droplets exist in the green domain.
In the same domain, in the case $\beta>0$, extensive droplets are found (now red domain). 
For $\beta>0$, droplets behave like passive ones only in the associative case of $\alpha = \pi/4$ (green line). 
Furthermore, an additional regime of intensive active droplets opens up (blue domain). This regime gets wider for higher values of the activity parameter $\beta$.
The boundary of extensive droplets is given by \eq{psi_crit}, and the outer boundary of intensive droplets is given by \eq{psi_bin}. For $\beta=0$, i.e., in the passive case, both bounds collapse.
Details on the analytical calculation of $\bar{\psi}^\pm_\text{crit}$
are given in Appendix~\ref{app:psi_crit}, and  parameter choices are listed in Appendix~\ref{sect:parameters_methods}.
}
\end{figure*}

The conserved quantity $\bar{\psi}$ (\eq{cons_quantity}) affects the non-equilibrium steady state and leads to a changed behavior of the stationary radius $R_\text{stat}$ as a function of the system size $R_\text{sys}$.
This changed behavior is depicted in \fig{system_size}(a) which shows  $R_\text{stat}(R_\text{sys})$ for two different values of the conserved quantity $\bar{\psi} = -0.61$ (blue) and $\bar{\psi}=-0.425$ (red) in a system with $\alpha=0$ and $\beta=\pi/8$. 
Interestingly, in the case of $\bar{\psi}= -0.61$ (blue), the stable stationary solution converges to a finite value in the limit of large systems, i.e., $R_\text{sys} \rightarrow \infty$, while for $\bar{\psi}= -0.425$ (red), the solution scales linearly with the system size, for large systems. In other words, there are two different cases: 
(i) \textit{Intensive} active droplets (blue) for which the stationary droplet size $R_\text{stat}$ quickly saturates with system size $R_\text{sys}$ implying that $R_\text{stat}$ 
is set by molecular and kinetic parameters for large system size. 
(ii) \textit{Extensive} active droplets (red) where the stationary droplet size $R_\text{stat}$ increases linearly with system size $R_\text{sys}$ implying that $R_\text{stat}$ is set by $R_\text{sys}$.

To explore these two behaviors, we numerically solved the dynamic equations with a continuous interface (Eqs.~\eqref{eq:dyn_phi}) in three dimensions. 
\fig{system_size}(b) shows the field $\phi_A(\vect{x})$ in four different stationary states, corresponding to the two different cases (i,ii) and two different system sizes ($L=600$ and $L=150$ in inset). 
We consider the same values of the conserved quantity as in  \fig{system_size}(a).
In the case of $\bar{\psi}=-0.61$ (blue), the radius of the stationary active droplet for the larger system is almost identical to that of the smaller system in the inset. 
In the case of $\bar{\psi}=-0.425$ (red), however, the stationary droplet radius in the larger system is much larger than in the smaller box. 
We conclude that the solutions of the continuous dynamic Eqs.~\eqref{eq:dyn_phi} confirm the trend obtained from our single droplet study in the sharp interface limit discussed in  \fig{system_size}(a). 

The qualitative change in the scaling of the stationary radius $R_\text{stat}$ with system size $R_\text{sys}$ 
upon changing the conserved quantity  $\bar{\psi}$ suggests that there is a bifurcation between the regimes of intensive (blue) and extensive (red) active droplets. 
To determine the nature of this bifurcation, we use the sharp interface model of a single droplet and consider the limit of an infinite system; details see Appendix~\ref{app:edm}.
We calculate the stable (solid) and unstable (dashed) stationary radii $R_\text{stat}$ as a function of the conserved quantity $\bar{\psi}$ (\fig{system_size}(c)). 
We find that $R_\text{stat}$ diverges at the threshold values (see Appendix~\ref{app:psi_crit} for details on the derivation)
\begin{align}
\label{eq:psi_crit}
\bar{\psi}_\text{crit}^\pm = \pm \frac{\phi_0}{2}  \Big( \cos(\alpha) + \sin(\alpha) \Big)
\qd
\end{align} 
In the vicinity of these critical values, the stationary droplet radius diverges with $R_\text{stat} \propto |\bar{\psi}-\bar{\psi}_\text{crit}^\pm|^{-1}$ within the blue domain, see App.~\ref{app:psi_crit}.

Furthermore, stationary active droplets can only be found within a certain range of the conserved quantity $\bar{\psi}\in [-\bar{\psi}_\text{bin},\bar{\psi}_\text{bin}]$ (dashed-dotted black line in \fig{system_size}(c)) with
\begin{gather}
\label{eq:psi_bin}
\bar{\psi}_\text{bin}^\pm = \pm \frac{\phi_0}{2}\frac{\cos(\alpha+\beta) + \sin(\alpha+\beta)}{\cos (\beta)} \qd
\end{gather}
This range in the conserved quantity graphically corresponds to steady states for which the reaction nullcline lies within the binodal domain of the phase diagram (\fig{ph_d}). 
Moreover, for $-\bar{\psi}_\text{bin}<\bar{\psi}<0$, we find $A$-rich droplets in a solvent-rich phase, while for $0<\bar{\psi}<\bar{\psi}_\text{bin}$, we find solvent-rich droplets in a $A$-rich phase.

By means of Eq.~\eqref{eq:psi_crit} and Eq.~\eqref{eq:psi_bin}, we can study how the underlying molecular interactions (parameterized by the compositional angle $\alpha$) and the activity parameter $\beta$ affect the existence of droplets and the critical bifurcation; see Fig.~\ref{fig:state_diag}(a,b).
Neglecting the finite size effects, 
these results are independent of the specific values of kinetic parameters such as $\Gamma$ and $K$, and otherwise only depend on thermodynamic parameters in the free energy (\eq{f}). 
We find that for $\beta>0$, the active driving leads to a regime of intensive droplets for all values $-\pi/4 \leq \alpha< \pi/4$, see Fig.~\ref{fig:state_diag}(a). Recall that in the segregative case, $\alpha = -\pi/4$ corresponds to an effective binary model of components $A$ and $B$, which phase separate from each other and are converted into each other via chemical reactions; this case was studied in Ref.~\cite{zwicker_suppression_2015}. Only in this limiting case are droplets always intensive, and the conserved quantity does not affect the stationary radii of chemically active droplets. For all values $-\pi/4<\alpha< \pi/4$, there is an additional domain with extensive droplets. The transition between these two regimes occurs at a value of the conserved quantity $\psicritpm$ expressed in Eq.~\eqref{eq:psi_crit}. 
In the associative case $\alpha=\pi/4$, $\psicritpm=\psibinpm$, i.e., no intensive droplets can be found, independently of the value of the activity parameter $\beta$. In this case, droplets behave like droplets in passive systems despite the presence of an active chemical reaction. Chemical reactions still drive diffusive fluxes of $A$ and $B$ between the phases. However, these fluxes do not affect the spatial distribution of the stationary conserved field  $\psi(x)$.
In the passive case with $\beta=0$, we find that $\bar{\psi}_\text{bin} = \bar{\psi}_\text{crit}$ independent of the compositional angle $\alpha$, see Fig.~\ref{fig:state_diag}(b) leftmost panel (green domain). Indeed, passive droplets are always extensive since they behave as thermodynamic phases that scale with the system size. 
However, for $\beta>0$, $\bar{\psi}_\text{crit}^\pm \lessgtr \psi_\text{bin}^\pm$. Therefore, intensive droplets of finite size exist inside a domain that increases for larger activity parameter $\beta$ (see Fig.~\ref{fig:state_diag}(b) from left to right).

In summary, our analysis of single active droplets shows that there are two different classes of stationary states in the limit of an infinitely large system: 
intensive active droplets and extensive active droplets. While an intensive active droplet adopts a finite size in a large system, an extensive active droplet scales with the system size. 
There is a critical transition between both types of non-equilibrium steady states. 
This transition is controlled by the interactions among the components, characterized in our model by the compositional angle $\alpha$, the conserved quantity $\bar \psi$, and the activity parameter $\beta$. 
In the next section, we  study emulsions composed of many active droplets and explore their behavior in the parameter regimes corresponding to intensive and extensive active droplets. 

\begin{figure*}
\centering
\includegraphics[width=0.99\textwidth]{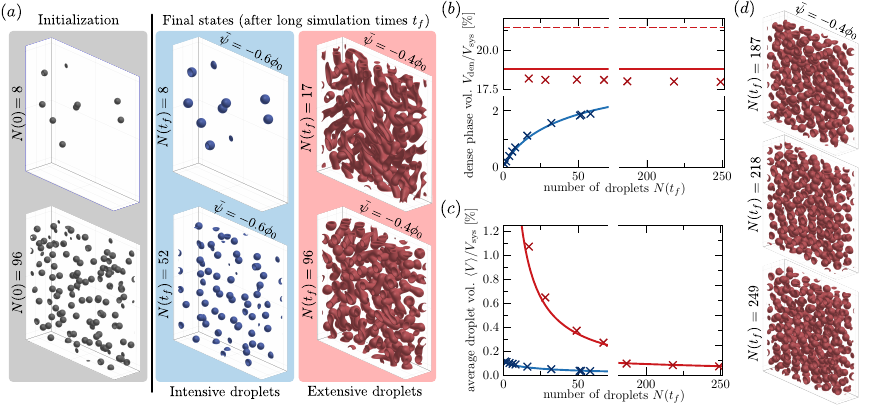}
\caption{\textbf{Stationary states in emulsions composed of many intensive (blue) and extensive (red) active droplets}: \textbf{(a)} 
Density plots of $\phi_A$ with $N(t)$ droplets for the initial conditions (gray) of $N(0) = 8$ and $N(0) = 96$ at time $t=0$ and the corresponding final states after long simulation times $t_f=5000$ (units $1/K$, Eq.~\eq{overall_reaction_rate}) for $\bar{\psi}=\phi_0$ (blue) and $\bar{\psi}=-0.4$ (red). 
\textbf{(b)} Numerically measured total volume of the dense phase $V_\text{den}=\langle V \rangle N$ as a function of the final droplet number $N(t_f)$  blue and red/blue points  ($\bar{\psi}=-0.6\, \phi_0 / \bar{\psi}=-0.4 \, \phi_0$) marked with cross symbols. Continuous lines show the results obtained by dividing the system volume into $N$ equal subvolumes and carrying out the single droplets analysis. The red dashed line shows the resulting fraction of the phase volume in the limit of infinitely fast diffusion~\cite{bauermann_chemical_2022}. 
\textbf{(c)} Average droplet volume $\langle V \rangle = N^{-1} \sum_i^N V_i $, where $V_i $ are individual droplet volumes. As for the upper panel, crosses correspond to numerical results, and lines correspond to the single droplet analysis.   
\textbf{(d)} Stationary concentration profiles $\phi_A$ for $\bar{\psi}=-0.4 \, \phi_0$ corresponding to a final droplet number of $N(t_f)=187$ ($N(0) = 192$), $N(t_f)=218$ ($N(0) = 256$), and $N(t_f)=249$ ($N(0) = 352$) from top to bottom.
In all cases, $\alpha = 0$, $\beta = \pi/8$, $K=0.0028$. After the final time $t_f=5000$, no significant changes were detectable for all cases.
Parameters used for solving the continuous model~\eq{dyn_phi} are given in Appendix \ref{sect:parameters_methods}.
} 
\label{fig:many_d}
\end{figure*}

\section{Interactions of many droplets} 
\label{sec:emulsion}

In this section, we study the dynamics in active emulsions by numerically solving the dynamic equations (\eq{dyn_phi}) for the continuous fields $\phi_A(\vect{x},t)$ and $\phi_B(\vect{x},t)$  with a continuous interface in three dimensions.
We initialize the system in a homogeneous state at local chemical equilibrium, with the average concentration values $\bar{\phi}_A$ and $\bar{\phi}_B$ inside the binodal domain but outside of the spinodal domain. 
Moreover,  $N_\text{init}$ small spherical droplets above the critical nucleation radius are randomly positioned in the system with $\phi_A=\phi_0$ inside (Eq.~\eqref{eq:f}). 
To avoid the fusion of such initially placed droplets right after initialization, we only accept random configurations with inter-droplet distances above a threshold value. 

We now discuss representative results corresponding to the intermediate values of the compositional angle $\alpha=0$ and the activity parameter $\beta=\pi/8$.
With this choice, \eq{psi_bin} and \eq{psi_crit} imply that the bounds on the conserved quantity $\bar \psi$ for the existence  of droplets are given by $\bar{\psi}_\text{bin}^\pm = \pm \phi_0/\sqrt{2}$ and the critical bifurcation between intensive and extensive droplets is $\bar{\psi}_\text{crit}^\pm = \pm 0.5 \,\phi_0$. 
The initial concentrations outside are chosen such that the total conserved quantity in the system is either $\bar{\psi}=-0.6 \,\phi_0$, or $\bar{\psi}=-0.4 \,\phi_0$. 
These two values lie, respectively, in the regime where chemically  active droplets are intensive ($\bar{\psi}<\bar{\psi}_\text{crit}^-$), or extensive ($\bar{\psi}_\text{crit}^-<\bar{\psi}<\bar{\psi}_\text{crit}^+$).

The left-most panels of \fig{many_d}(a) (grey) depicts the initial state of the active emulsion  corresponding to an initial droplet number of $N_\text{init} = 8$  and $N_\text{init} = 96$, respectively. 
In the remaining columns of \fig{many_d}(a) (blue and red),
we show the
the stationary states of the active emulsion for these two initial conditions and the aforementioned conserved quantities  $\bar{\psi}=-0.6 \,\phi_0$ (blue) and $\bar{\psi}=-0.4 \,\phi_0$ (red). 
We find that for $\bar{\psi}=-0.6 \,  \phi_0$ (intensive regime), all droplets grow to the same size, i.e., a mono-disperse active emulsion emerges. 
During this kinetics to the stationary state, a few droplets vanish compared to the initial condition. 
Interestingly, for  $\bar{\psi}=-0.6 \,  \phi_0$, the stationary droplet radius changes only slightly for the two initial conditions. 
In the right column of \fig{many_d}(a) (red), we show the stationary states for $\psi=-0.4 \, \phi_0$ (above the critical transition, extensive regime) for the same initial conditions as above. For the initial state composed of eight droplets only (first row), droplets elongate and sometimes branch, thereby forming tube-like structures. However, some of these branched structures can separate during the dynamics such that, in the end, $N=17$ different tubes of different lengths exist. 
When initializing $96$ droplets (second row), some of them elongate, while others stay almost spherical, dependent on the initial distance to other droplets. However, none of them dissolve, such that we find $N=96$ objects in the stationary state. When even more droplets are initialized, some of the initial droplets dissolve, while the rest stay spherical and try to maximize their distance towards each other, see  \fig{many_d}(d).

Despite these complex changes in droplet morphologies, there is a common principle in the regime where single active droplets are  extensive (red): 
Despite varying the initial conditions and different numbers of droplet-like domains in the stationary state, the total phase volume is almost constant (\fig{many_d}(b)). 
In other words, when there are more droplet-like domains, they are smaller such that the total phase volume is approximately conserved. 
In contrast, in the regime where active droplets are intensive (blue), the total volume increases with the  amount of initialized droplets. Most importantly, in this regime, the average droplet volume is roughly constant (\fig{many_d}(c)) and approximately equal to the  volume of one single droplet up to finite size effects.

Monodisperisty in chemically active emulsions can be explained as follows:
Intensive droplets (blue) adopt a fixed size independent of the system size. When just a few droplets are initialized in a large system, they can be far apart, and whenever they are separated by a distance much longer than the reaction-diffusion length scale in the outside phase, they hardly interact with each other. As a consequence, their dynamics becomes stationary and their average radius takes a value similar to the stationary radius $R_\text{stat}$ obtained from the single droplet analysis (see Sect.~\ref{sect:single_drop}).
When more droplets are initialized, such that they are closer than this reaction-diffusion length scale, they weakly interact and get stationary at smaller sizes. This leads to the slight decrease of average droplet volume for large $N$, seen in \fig{many_d}(c).
In all cases, droplets reach identical radii.
For extensive droplets (red), monodispersity is only found when many droplets are initialized. Indeed, when only a few droplets are initialized, they tend to grow until they feel the presence of their neighbors. The more droplets are initialized, the sooner this arrest occurs, decreasing the radii of droplets. 
If the number of initialized droplets is too small, droplets grow to sizes where shape instabilities can occur. 
Tubes form via an elongation instability, discussed in the next section, or spherical shells form via a spinodal instability at the center of large droplets~\cite{bauermann_formation_2023}. 
Crucially, when enough droplets are present, the arrest of growth happens earlier than shape instabilities can occur. As a consequence, droplets are spherical and reach identical radii (\fig{many_d}(d)).
In summary, for extensive droplets in an emulsion (red),
the history of the emulsion matters for the final stationary mono-dispersed state or the emerging shape instabilities.

To confirm the validity of these arguments relying on a single droplet analysis,  we compare the phase volume and the average droplet size with the analytic results obtained from the single droplet case in the sharp interface limit.
To this end, we considered $N$ identical subsystems arranged in a hexagonal close-packed lattice that fills the total system volume $V_\text{sys}$, and calculated phase volumes and average sizes. 
These analytically derived results (solid lines in \fig{many_d}(b,c)) are in good agreement with the measured values in the numerical simulations of the active emulsions. 
In the case of extensive droplets (dashed lines in \fig{many_d}(b,c)), the total volumes of the phases are well captured by considering the limit of fast diffusion~\cite{bauermann_chemical_2022}. 

\begin{figure}
\centering
\includegraphics[width=0.45\textwidth]{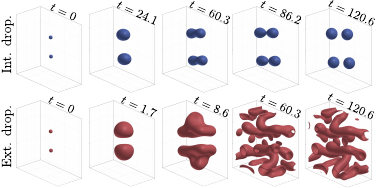}
\caption{\label{fig:division} \textbf{Time evolution of shape unstable intensive (blue) and extensive (red) active droplets}:
Snapshots of concentration profile $\phi_A$ for a conserved quantity $\bar{\psi}=-0.6\,\phi_0$ (blue/upper row) and $\bar{\psi}=-0.4\,\phi_0$ (red/lower row).
While intensive droplets divide, extensive droplets undergo a Mullins-Sekerka instability and form complex tube-like morphologies.
Parameter choices see Appendix~\ref{sect:parameters_methods}.}
\end{figure}

\section{Shape instabilities and droplet division}
\begin{figure*}
\centering
\includegraphics[width=0.9\textwidth]{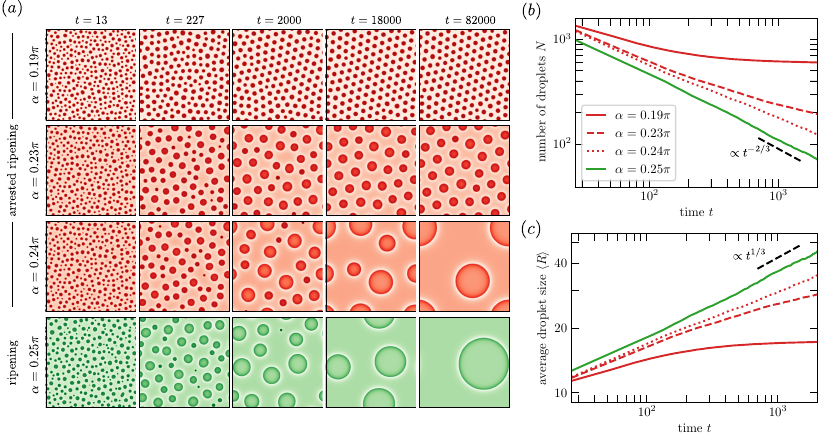}
\caption{\label{fig:ripe}\textbf{Arrested ripening of extensive active droplets (red)}: \textbf{(a)} Snapshots of concentration field $\phi_A$ evolving in time, for an activity parameter $\beta=\pi/4$
and
four different values of the compositional angle  ($\alpha = 0.19 \pi,\; 0.23 \pi,\; 0.24\pi,\; 0.25\pi$ from top to bottom), each corresponding to different molecular interactions. All simulations are initialized using the concentration fields shortly after the spinodal instability of the homogeneous chemical steady state with $\bar{\psi}=-0.35\phi_0$ has occurred.
\textbf{(b,c)}  Droplet number $N$ and average droplet radius $\langle R \rangle$ as a function of time, for the same values of $\alpha$. Passive scaling laws are indicated with dashed black lines.
Parameters are given in Appendix ~\ref{sect:parameters_methods}.
} 
\end{figure*}

In the previous section, we have seen that for extensive active droplets in an emulsion, an elongation instability can occur. 
This instability is reminiscent of the Mullins-Sekerka instability that occurs in the diffusive growth of a single droplet without reactions in a large system. 
It requires a (weak) deformation of a droplet's spherical shape and relies on the material depositing at interfacial domains of larger mean curvature.
As a result, these domains grow faster, further deforming the droplet shape. 
Deformations are counteracted by surface tension, which tends to flatten the mean curvatures of the interface. However, this  effect weakens the larger the droplet becomes. Thus, there is a critical radius above which droplets constant growth and deform (Mullins-Sekerka instability).

A similar instability can occur for both extensive and intensive active droplets.
Indeed, in a binary mixture where droplets are always intensive, such instability was shown to lead to droplet division~\cite{zwicker_growth_2017}.
Here, we show that droplet dynamics after elongation vastly differ between intensive and extensive active droplets.
In the upper row of \fig{division} (blue), we show snapshots of the dynamics of two intensive droplets ($\bar{\psi}=-0.6\, \phi_0)$ and compare it to the dynamics of two extensive droplets ($\bar{\psi}=-0.4\, \phi_0)$ in the lower row (red). For both cases, we choose identical kinetic parameters.
However, we have chosen the overall reaction rate $K$ (\eq{overall_reaction_rate}) such that the stationary droplet radius in the intensive regime exceeds the critical radius for onset of the elongation instability.
In both cases, the two droplets are initially separated by a distance smaller than the reaction-diffusion length in the outside phase. Thus, the presence of the neighboring droplets leads to asymmetries that induce initial deformations, which get amplified by the elongation instability.
However, the division predominantly happens for intensive droplets; see \fig{division} (upper row). Extensive droplets elongate (\fig{division} lower row) and form tubes, as seen before.
This difference can be explained by focusing on the neck region. For intensive droplets, there is almost no growth in this region, such that a Plateau-Rayleigh instability can lead to a pinch-off. However, extensive droplets constantly grow, including in the neck region, thereby inhibiting the Plateau-Rayleigh instability.

Interestingly, with such cycles of growth and division of intensive droplets, the total phase volume can trivially grow just by increasing the number of droplets. As a result, these shape instabilities allow intensive droplets to fill the space. Thus, macroscopic phase volumes are reached, even in infinite systems, similar to the regime of extensive droplets.

\section{Molecular interactions affect the arrest of ripening}
\label{sect:arrest_ripe}

In section~\ref{sec:emulsion}, we have shown that extensive droplets grow until the interaction with neighboring droplets arrests their growth. However, in section~\ref{sec:dyn_con_noncon}, we discussed that in the associative case ($\alpha = \pi/4$) where $A$ and $B$ have identical interactions with the solvent, the conserved field decouples from the active driving, leading to passive phase-separating dynamics. 
In this section, we study how the emulsion dynamics is affected as molecular interactions approach the limit of the associative case ($\alpha \to \pi/4$). We find that extensive droplets can initially ripen like in the passive case until the effects of the active driving manifest and ripening arrests.  
 
In \fig{ripe}(a), we show snapshots of the dynamics of concentration field $\phi_A$ for four different values of the compositional angle $\alpha$; see Supplementary Material for videos on the ripening dynamics. All numerical calculations are conducted with the same kinetic parameters, an activity parameter $\beta=\pi/4$, and a conserved quantity of $\bar{\psi} = -0.35 \phi_0$. For these settings, all systems are populated by extensive droplets.
We initialize at $t=0$ (not shown in the figure), the concentrations homogeneously at chemical steady state with small fluctuations (around $0.1\%$ of $\bar{\psi}$). 
For the values of the compositional angle $\alpha$ considered, the system is within the spinodal regime of a non-reacting phase separating mixtures. For triggering the nucleation of many droplets, seen in the first row in \fig{ripe}, we did not allow for chemical reactions in the early times of the spinodal instability ($t\approx 2$). Insights on the change of the nucleation dynamics in chemically active systems can be found here~\cite{Cates2023, Ziethen2023}.
While the nucleation process is similar for all the compositional angles $\alpha$ considered here, the long-time ripening dynamics depend strongly on $\alpha$. 
For $\alpha=0.19 \, \pi$ (red/first row in \fig{ripe}(a)), some of the initial droplets dissolve, but the interaction between droplets causes the ripening to arrest.
As the interaction of $A$ and $B$ with the solvent gets more similar, i.e., $\alpha \to \pi/4$ (associative case), this arrest of ripening occurs at a later stage of the dynamics. Prior to the arrest of ripening, smaller droplets dissolve while larger ones grow. Thus, as the arrest occurs and the emulsion becomes monodisperse, stationary droplet radii are larger as $\alpha \to \pi/4$ (associative case), see second ($\alpha = 0.23 \pi$) and third ($\alpha = 0.24 \pi$) row of \fig{ripe}(a), in red. Only in the associative case  $\alpha = \pi/4$, the arrest is absent and chemically active droplets ripen as passive emulsions (green/last row in \fig{ripe}(a)). This means that the classical power-laws of Ostwald ripening hold in the associative case ($\alpha = \pi/4$) with $\langle R \rangle \propto t^{1/3}$ and $N\propto t^{-2/3}$~\cite{Krichevsky1993},
while for decreasing values of $\alpha < \pi/4$, the transition to such classical ripening laws happens at later stages of the dynamics (see \fig{ripe}(b,c)).

\section{Discussion}

We have introduced and studied a model of a minimal ternary mixture of the components $A$, $B$, and solvent that can phase separate while $A$ and $B$ convert into each other. 
This conversion is maintained away from chemical equilibrium 
by choosing reaction fluxes independent of the chemical potentials. 
This choice prevents the system from simultaneously reaching chemical and phase equilibrium and allows it to adopt a non-equilibrium steady state.
The model is minimal because it exhibits a non-trivial conserved quantity of the reaction with an associated conserved field that is generally position-dependent, also in the non-equilibrium steady state. 
Within this minimal model for active chemical reactions in phase-separating mixtures, we found a critical transition, determining whether the size of a single droplet scales with system size (extensive active droplet), like in passive systems,
or whether it reaches a finite, characteristic size (intensive active droplet). 

The transition found in the single droplet analysis has crucial consequences for the behavior of emulsions composed of many droplets. For a value of the conserved quantity in the regime of an intensive droplet, a mono-disperse emulsion forms where the average size is approximately given by the stationary size of a single droplet. The reason for this approximate agreement is that  droplets interact only weakly in the intensive droplet regime. 
In contrast, extensive droplets in an emulsion constantly grow until they come close enough to interact strongly with each other. Once they interact, there can be initial ripening, but growth arrests. 
The morphology of the resulting structures depends crucially on initial conditions, i.e., the history of the emulsion.  In other words, few droplets deform via shape instabilities, while many droplets remain spherical and monodisperse. However, the  total phase volume is independent of the number and sizes of droplets and scales with the system volume similar to passive thermodynamic phases.

Our findings highlight the importance of several components in chemically active systems.
For example, accounting for a solvent component fundamentally alters the dynamics of chemically active droplets, reflected in the  appearance of the critical transition when varying the conserved quantity. 
We propose that understanding the effects of the conserved quantities(s) in reacting system away from equilibrium is crucial to correctly interpret emulsion dynamics in biological and chemical systems~\cite{brangwynneGermlineGranulesAre2009, fericCoexistingLiquidPhases2016, shinLiquidPhaseCondensation2017, boeynaemsProteinPhaseSeparation2018, sastre2024size}.
Specifically, our theory can unravel the conserved quantity's effects in synthetic chemical systems, such as droplet size control in multi-component mixtures leading to chemically active coacervates~\cite{sastre2024size}. 
According to our theory, droplet division is more likely to occur in  the regime $\alpha \simeq \pi/4$  corresponding to intensive active droplets and the case where $A$ segregates from $B$ from each other (segregative case). 
We propose  studying a single active droplet in a reaction container for different container sizes~\cite{Donau2020} and test whether or not
the active droplet scales with the container size (intensive active droplets, see Fig.~\ref{fig:system_size}(a)).
In the regime of intensive active droplets, division is more likely to be obtained.  This procedure could be used to test synthesized actively reacting components for their division propensity and may thus pave the way to observe successive division events in larger, mass-supplied containers containing many active droplets. 

\bibliography{lib}
\newpage

\section*{Appendix}
\appendix
\section{Chemical reaction rates}
\label{app:chem_rates}

In the following, we will use the overall rate $K = k_{AB} + k_{BA} $ and the activity parameter $\beta$ as parameters for the chemical reactions. These parameters are related to the forward rate $k_{BA}$ and backward rate $k_{AB}$, appearing in Eqs. \eqref{eq:dyn_phiA} via the following relationships: $k_{BA} = K \sin(\alpha+\beta)/[\cos(\alpha+\beta)+ \sin(\alpha+\beta)]$ and $k_{AB} = K \cos(\alpha+\beta)/[\cos(\alpha+\beta)+ \sin(\alpha+\beta)]$, where $\alpha$ is the compositional angle. Using this parameterization, the contour lines of $\bar \psi$, where the conserved quantity is constant, have a fixed slope of $-1$ that is
independent of the compositional angle $\alpha$.  However, note that the slope of the chemical steady state changes when varying $\alpha$. Moreover, changes in $\alpha$ do not affect the activity in the system, in the sense that they do not alter the angle between the reaction nullcline and the tie lines. Thus, with this new parameterization, we can vary the compositional angle $\alpha$ for studying how different interactions affect active droplets while keeping the activity parameter $\beta$ fixed. The rate $K$ sets only a time scale.

\section{The sharp-interface limit}
\label{app:edm}

Similar to the droplet dynamics of passive phases (see Ref.~\cite{brayTheoryPhaseorderingKinetics1994} for example), we can study the dynamics of a single active droplet in the limit of a sharp interface. 
This approach for general multicomponent mixtures is discussed in detail in Ref.~\cite{bauermann_energy_2022}. In this section, we derive the stationary state conditions of a chemically active droplet in the sharp-interface limit for a ternary mixture described by the free energy given in Eqs.~\eqref{eq:f}. With this choice of free energy, analytical results for the stationary radius and the critical value for the transition between intensive and extensive active droplets can be determined.

For a single droplet with radial symmetry sitting at the origin ($r$ being the radial coordinate), we split the field into two domains (inside/outside, abbreviated in/out in the following) and couple these domains via boundary conditions at an interface at position $r=R$. 
Furthermore, we linearize the dynamics of the fields $\phi_A$ and $\phi_B$ in \eq{dyn_phi} around the corresponding $\phi$-values at the interface. 
In this approximation, the stationary state of a chemically active droplet in our ternary mixture is determined by the $\phi$-values of $A/B$ on both sides of the interface, named $\Phi_A^\ins$, $\Phi_A^\out$, $\Phi_B^\ins$, and $\Phi_B^\out$, and the interface position $R$. In the following, we explain how these five values are determined.

Locally, at the interface, we assume an equilibrium of phase separation, i.e., identical chemical potentials and osmotic pressure across the interface.
Furthermore, global conservation laws and vanishing flux differences across the interface fix a unique phase equilibrium locally at the interface and, therefore, a unique non-equilibrium steady state of the droplet, see Ref.~\cite{bauermann_energy_2022}.

For our simple free-energy model~\eqref{eq:f}, we can obtain the four values of $\Phi_i^\text{in/out}$ generally as a function of $R$ and $\alpha$.
The determination of the stationary position of the interface $R_\text{stat}$  additionally needs flux relations (see below). Assuming local phase equilibrium at the interface constrains its concentrations to the two straight binodal lines shown in \fig{ph_d}(b) for an infinitely large system. Laplace pressure effects of an interface with surface tension $\gamma$ of finite-sized droplets are corrected up to linear order via the so-called Gibbs-Thomsom coefficients $c$~\cite{zwicker_suppression_2015}. We parameterize this phase equilibrium by the intersection of its tie line with the  $\phi_2$-axis in its $\phi_1$ and $\phi_2$ representation (as defined in \eq{f}). This intersection we call $\phi_2^\text{inter}$, and write
\begin{align}
\label{eq:interface_cond_app}
\begin{pmatrix}
\Phi^\text{in/out}_A\\
\Phi^\text{in/out}_B
\end{pmatrix} 
&=
\begin{pmatrix}
\cos \alpha & \sin \alpha\\
-\sin \alpha & \cos \alpha
\end{pmatrix} 
\begin{pmatrix}
\pm \z + \gamma c H\\
\phi_2^\text{inter}
\end{pmatrix} \qc
\end{align}
where we assumed that the conserved quantity is enriched inside and diluted outside. For such a spherical symmetric droplet, the mean curvature $H$ is given by $H = 2/R$. If the dilute phase builds up the spherical droplet, the two labels in/out in \eq{interface_cond_app} have to be swapped and $H=-2/R$.
Finally, we have to determine the intersection $\phi_2^\text{inter}$ via a global conservation law of the conserved quantity. Therefore, we note that while $\phi_A$ and $\phi_B$ have gradients in space, the conserved quantity is constant in space but jumps at the interface.
Thus, for a finite radial-symmetric system (system size $R_\text{sys}$) with an average amount of the conserved quantity $\bar{\psi}$, we know
\begin{equation}
 	\frac{1}{2}(\Phi^\ins_A+\Phi^\ins_B) R^3 + \frac{1}{2}(\Phi^\out_A+\Phi^\out_B) (R_\text{sys}^3 - R^3) = \bar{\psi} R_\text{sys}^3 \qd
\end{equation}
Using \eq{interface_cond_app}, we find
\begin{align}
\label{eq:Bintersec_finit_app}
	&\phi_2^\text{inter} = \nonumber \\  &\frac{2 R R_\text{sys}^3 \bar{\psi} + (2 \z R^4 - (2 c \gamma+ \z R) R_\text{sys}^3) (\cos(\alpha) + \sin(\alpha))}{R R_\text{sys}^3 (\cos(\alpha) - \sin(\alpha))}.
\end{align}
In an infinite system, however, the finite-sized droplet does not contribute to the average. Thus, the outside concentrations must fulfill
 \begin{equation}
   \bar{\psi} = \frac{1}{2}(\Phi^\out_A+\Phi^\out_B) \qc
 \end{equation}
and  therefore, again by using \eq{interface_cond_app}, we find
\begin{equation}
\label{eq:Bintersec_infinit_app}
   \phi_2^\text{inter} = \frac{2 R \bar{\psi}  - (2 c \gamma + \z R) (\cos(\alpha) + \sin(\alpha))}{R (\cos(\alpha) - \sin(\alpha))} \qd
 \end{equation}
We note by passing that it is only possible to derive the interface concentrations independently of the diffusivities and kinetic reaction rates due to our very symmetric form of the free energy density. Thus, these kinetic coefficients have to be taken into account only for the derivation of the stationary radius. In general, all five values have to be determined in parallel, a step that typically requires numerical solving schemes, see Ref.~\cite{bauermann_energy_2022}.
    
An interface can only be stationary when the diffusive fluxes $j_i^\text{in/out}$ are balanced across the interface, i.e. $j_A^\ins(R)= j_A^\out(R)$, and $j_B^\ins(R)= j_B^\out(R)$. 
These diffusive fluxes in the stationary state arise from the stationary concentration profiles $\phi_i^\text{in/out}(r)$ via $ j_i^\text{in/out}= - D_i^\text{in/out} \partial_r \phi_i^\text{in/out}(r)$, where $D_i^\text{in/out}$ is the diffusion coefficient of component $i$ in the corresponding domain obtained by linearizing. By construction, it is guaranteed that $j_A^\out = - j_B^\out$ and $j_A^\ins = - j_B^\ins$, such that once one of the flux balances is fulfilled, the second follows trivially.
Therefore, we derive only $j_A^\text{in/out}$ in the following.

After linearizing, we can determine these fluxes through the analytical stationary $\phi_A$-profiles in the two domains (solving the corresponding inhomogeneous Laplace equation in radial symmetry). From here, we can derive the total flux profiles, including the fluxes at the interface. We find
\begin{align}
\label{eq:jAin_app}
    j_A^\ins(R) &=  \Big(\Phi_B^\ins(R) k_{AB}^\ins - \Phi_A^\ins(R) k_{BA}^\ins\Big)
    \nonumber \\
    &\times \left( \frac{\coth(\lambda^\ins R)}{\lambda^\ins} - \frac{1}{(\lambda^\ins)^2 R} \right) \qc
\end{align}
where the rates $k_{ij}^\ins$ are the linrearized reaction rates from \eq{dyn_phi} and $\lambda^\ins = \sqrt{(D_A^\ins k_{AB}^\ins + D_B^\ins k_{BA}^\ins)/(D_A^\ins D_B^\ins)}$. Please note, that we use the dependency of the interface concentration $\Phi_i^\ins(R)$ on the position of the interface $R$ derived above.
However, in the outside domain ($r>R$), the solutions read
\begin{widetext}
\begin{align}
\label{eq:jAout_finite_app}
    j_A(R)^\out(R) &= \Big(\Phi_B^\out(R) k_{AB}^\out - \Phi_A^\out(R) k_{BA}^\out\Big)\nonumber\\
&\times \frac{ \lambda^\out (R - R_\text{sys}) \cosh(\lambda^\out (R - R_\text{sys})) + ((\lambda^\out)^2 R R_\text{sys}-1) \sinh(\lambda^\out (R - R_\text{sys}))
    }{(\lambda^\out)^2 R
    (\lambda^\out R_\text{sys} \cosh(\lambda^\out (R - R_\text{sys})) + \sinh(\lambda^\out (R - R_\text{sys})))} \qc
\end{align}
\end{widetext}
for a finite radial symmetric system with a no-flux boundary condition at the system size $R_\text{sys}$, or 
\begin{align}
\label{eq:jAout_infinite_app}
    j_A(R)^\out(R)  &= -\Big(\Phi_B^\out(R) k_{AB}^\out - \Phi_A^\out(R) k_{BA}^\out\Big) \nonumber \\
    &\times
    \frac{ k_{BA}^\out + \frac{D_A^\out}{D_B^\out} k_{AB}^\out +  (k_{AB}^\out + k_{BA}^\out) \lambda^\out R
    }
    {(k_{AB}^\out + k_{BA}^\out)(\lambda^\out)^2 R}
\end{align}
for infinite large systems, where again $k_{ij}^\out$ are the linrearized reaction rates from \eq{dyn_phi} and $\lambda^\out = \sqrt{(D_A^\out k_{AB}^\out + D_B^\out k_{BA}^\out)/(D_A^\out D_B^\out)}$.
The stationary droplet radii can now be found by using the interface concentrations stated in \eq{interface_cond_app} and \eq{Bintersec_finit_app} or \eq{Bintersec_infinit_app}, and numerically searching for the positions $R$ at which \eq{jAin_app} equals \eq{jAout_finite_app} or \eq{jAout_infinite_app}, depending on the system size.

\section{The critical value $\psi_{crit}$ and the scaling of the stationary radius in its vicinity}
\label{app:psi_crit}

In App.~\ref{app:edm}, we explained how the the stationary droplet radii can be obtained in the sharp interface limit. Here, we argue first how we can obtain the value of the conserved quantity at the transition in an infinite system and, second, show how the stationary radius scales in the vicinity of this transition.

To find the critical value of the conserved quantity, we balance the fluxes across the interface for an infinitely large droplet,
\begin{equation}
\label{eq:crit_trans_con_app}
\lim_{R\rightarrow \infty} j_A^\text{in}(R) = \lim_{R\rightarrow \infty} j_A^\text{out}(R) \qc
\end{equation}
from \eq{jAin_app} and \eq{jAout_infinite_app}. When \eq{interface_cond_app} and \eq{Bintersec_infinit_app} are applied in these expressions, a dependency on the average conserved quantity $\bar{\psi}$ follows. 
We find that there is only one value of this conserved quantity for which \eq{crit_trans_con_app} is true. This value reads
\begin{widetext}
\begin{align}
\bar{\psi}_\text{crit} =
-\frac{\phi_0}{2} 
\frac{ (D_A^\ins k_{AB}^\ins + D_B^\ins k_{BA}^\ins) (k_{AB}^\out + k_{BA}^\out) + D_A^\ins D_B^\ins \lambda^\ins \lambda^\out 
((k_{AB}^\ins - k_{BA}^\ins) \cos(2\alpha) + (k_{AB}^\ins + k_{BA}^\ins) \sin(2\alpha)
            )}
{\text{sin}(\alpha)\left(
(D_A^\ins k_{AB}^\ins  + D_B^\ins k_{BA}^\ins) (k_{BA}^\out + k_{AB}^\out\cot(\alpha))
+ D_A^\ins D_B^\ins  \lambda^\ins \lambda^\out  
(k_{BA}^\ins 
+ k_{AB}^\ins \cot(\alpha))
            \right)} \qd
\end{align}
\end{widetext}
When we apply our  parameterization of the reaction rates, i.e.,   
$k_{AB}^\text{in/out} = K^\text{in/out} \cos(\alpha+\beta)/[\cos(\alpha+\beta)+ \sin(\alpha+\beta)]$ and $k_{BA}^\text{in/out}= K^\text{in/out} \sin(\alpha+\beta)/[\cos(\alpha+\beta)+ \sin(\alpha+\beta)]$, see App.~\ref{app:chem_rates}, we find 
$\bar{\psi}_\text{crit}$ as a function of the kinetic rates $K^\text{in/out}$, diffusivities $D_i^\text{in/out}$, and the angles $\alpha$ and $\beta$. With the assumptions $D_A^\ins = D_A^\out$, $D_B^\ins = D_B^\out$ and $K^\ins= K^\out $, this expression simplifies and is solution of  \eq{psi_crit}. 
Furthermore, we considered a droplet of the dense phase in a dilute environment, which is the case for $\bar{\psi}<0$ ($-$ branch). For $\bar{\psi}>0$, the interface concentrations  swapped resulting in $+$ branch of the solution of \eq{psi_crit}.

Furthermore, we can check the scaling of the stationary radius of intensive active droplets in the vicinity of the critical transition. For this, we expand $ j_A^\out (R)= j_A^\ins (R) $ for large $ R$. We find the stationary radius $ R_\text{stat}$ as a function of $ \bar{\psi}$ that scales like $R_\text{stat} \propto |\bar{\psi} - \bar{\psi}_\text{crit}|^{-1}$. The general solution can be obtained from the equations above straightforwardly. Due to its length, however, in this appendix we restrict ourselves to the special case of $ D_A^\ins = D_A^\out = D_B^\ins = D_B^\out = D$ and $ K^\ins = K^\out =K $. We find for the case of a droplet of the dense phase in a dilute environment ($\bar{\psi}<\bar{\psi}_\text{crit}<0$)
\begin{equation}
\label{eq:analytics_Rstat}
    R_\text{stat} = -  \Lambda \phi_0 \tan(\beta) \frac{ 
     \cos(\alpha) - \sin(\alpha)
    }{\bar{\psi}- \bar{\psi}_\text{crit}} 
    \qc
\end{equation}
where we defined the length scale $\Lambda = \sqrt{D/K}$ and used the fact that for large droplets, the Laplace pressure becomes negligible.

\section{Parameter choices and methods used for figures}\label{sect:parameters_methods}

To numerically solve the dynamics equations in this work, we rescale time $t \cdot K$ with $K$ denoting the  overall rate (\eq{overall_reaction_rate}), 
and position $\vect{x}/\ell $ where $\ell=\sqrt{\phi_0^2 \kappa_A/b_1}$, thus, half of the interface width in the continuous model. 
For simplicity, we consider the overall rate equal inside and outside for all studies ($K^\text{in}=K^\text{out}=K$). 
These rescalings yield the following non-dimensional parameters in our numerical studies:
$D_i^\text{in/out} /(\ell^2K)$, $\gamma c/ \ell$, and $\Gamma/(\ell^2K b_1 )$.

\subsection{Figures showing results obtained in the sharp interface model}

In the previous sections, we used the sharp interface limit to calculate the results shown in 
Fig.~\ref{fig:system_size}(a,b).
Using a standard root-finding scheme, we can numerically solve the specific values of the stationary interfaces shown in Fig.~\ref{fig:system_size} in the main text. The parameter values used to produce these figures are $D_A^\io/(\ell^2K)=0.0014^{-1}$, $D_B^\io/(\ell^2K)=0.0014^{-1}$, $\alpha = 0$, $\beta = 0.12 \pi $, $\gamma c/\ell = 1/6$,  and $\phi_0 = 1$.
 
\subsection{Figures showing results obtained in the continuous model}
All the results obtained in the continuous model were obtained by using an implicit-explicit Runge-Kutta solver of the second-third order. We considered periodic boundary conditions and approximated the higher-order derivatives using pseudo-spectral methods on a regular lattice.
For simplicity, we used $b_1/\phi_0^2=b_2=\phi_0=1$, $\alpha = 0$, $\kappa_A = \kappa_B$ and $\beta=0.12\pi$ were not otherwise explicitly mentioned in the figure caption.
The mobilities were set to $\Gamma/(\ell^2K b_1)=0.0014^{-1}$ in Fig.~\ref{fig:system_size}, $\Gamma/(\ell^2K b_1 )=0.0028^{-1}$ in Fig.~\ref{fig:many_d}, $\Gamma/(\ell^2K b_1 )=0.00112^{-1} $ in Fig.~\ref{fig:division}, and  $\Gamma/(\ell^2K b_1 \phi_0^2)=0.004^{-1}$ in Fig.~\ref{fig:ripe}.
Furthermore, we used $N_1 \times N_2$ or $N_1 \times N_2 \times N_3$ grid points for two-dimensional or three-dimensional systems of sizes of $L_1 \times L_2$ or $L_1 \times L_2  \times L_3$, respectively.
The specific values for the corresponding figures are listed in Table~\ref{tab:parameters_cont}.
\begin{table}[h]
\label{tab:parameters_cont}
\begin{tabular}
{c|c|c}
            &  Grid & Size \\ \hline
Fig.~\ref{fig:system_size}(b) large   &  1024$\times$1024  & 600$\times$600      \\
Fig.~\ref{fig:system_size}(b) small  &   256$\times$256    & 150$\times$150      \\
Fig.~\ref{fig:many_d} &  512$\times$512$\times$128& 500$\times$500$\times$125      \\
Fig.~\ref{fig:division} &56$\times$256$\times$128& 250$\times$250$\times$125       \\
Fig.~\ref{fig:ripe}     &   512$\times$512    &       550$\times$550
\end{tabular}
\caption{Parameter for grid and system size used for figures showing numerical results obtained from solving the continuous model (\eq{dyn_phi}). }
\end{table}

\end{document}